\documentclass[journal]{vgtc}                     
\usepackage{amssymb,amsthm,enumitem}
\usepackage{amsmath}
\usepackage{mathtools}
\usepackage{amsfonts}
\usepackage{verbatim}
\usepackage{placeins}
\usepackage{comment}
\usepackage{algorithm2e}
\usepackage{algpseudocode}
\PassOptionsToPackage{warn}{textcomp}  
\usepackage{textcomp}                  
\usepackage{mathptmx}                  
\usepackage{times}                     
\usepackage{cite}                      
\usepackage{tabu}                      
\usepackage{booktabs}                  
\usepackage{adjustbox}
\usepackage{bm}
\usepackage{colortbl}
\usepackage{microtype}                 
\usepackage{siunitx}

\SetKw{Continue}{continue}

\usepackage{graphicx}
\usepackage{xfrac}
\usepackage{relsize}
\usepackage{soul}
\usepackage{color}
\usepackage{lipsum}

\newcommand {\mm}[1] {\ifmmode{#1}\else{\mbox{\(#1\)}}\fi}
\newcommand{\Mspace}{\mm{{\mathbb M}}}

\newcommand{\Rspace}{\mm{{\mathbb R}}}
\newcommand{\grad}[1]{{\nabla {#1}}}
\newcommand*\diff{\mathop{}\!\mathrm{d}}
\newlength\mylen

\newcommand\blfootnote[1]{%
  \begingroup
  \renewcommand\thefootnote{}\footnote{#1}%
  \addtocounter{footnote}{-1}%
  \endgroup
}

\onlineid{1277}

\vgtccategory{Research}

\vgtcpapertype{algorithm/technique}

\title{Uncertainty Visualization of Critical Points of 2D Scalar Fields for Parametric and Nonparametric Probabilistic Models}

\author{%
  \authororcid{Tushar M. Athawale}{0000-0003-3163-6274},
   \authororcid{Zhe Wang}{0000-0003-1123-9925}, \authororcid{David Pugmire}{0000-0002-7051-3288}, \authororcid{Kenneth Moreland}{0000-0002-7051-3288}, \authororcid{Qian Gong}{0000-0002-3570-4142}, \authororcid{Scott Klasky}{0000-0003-3559-5772}, \\ \authororcid{Chris R. Johnson}{0000-0001-5673-5338}, and \authororcid{Paul Rosen}{0000-0002-0873-9518}
}

\authorfooter{
  \item
  	Tushar M. Athawale, Zhe Wang, Qian Gong, Scott Klasky, Kenneth Moreland, and David Pugmire are with the Oak Ridge National Laboratory.
  	E-mail:\{athawaletm, wangz, pugmire, morelandkd, gongq, klasky \}@ornl.gov
  \item
  	Chris R. Johnson and Paul Rosen are with the University of Utah
  	E-mail: \{crj, prosen\}@sci.utah.edu
}

\abstract{%
  This paper presents a novel end-to-end framework for closed-form computation and visualization of critical point uncertainty in 2D uncertain scalar fields. Critical points are fundamental topological descriptors used in the visualization and analysis of scalar fields. The uncertainty inherent in data (e.g., observational and experimental data, approximations in simulations, and compression), however, creates uncertainty regarding critical point positions. Uncertainty in critical point positions, therefore, cannot be ignored, given their impact on downstream data analysis tasks. In this work, we study uncertainty in critical points as a function of uncertainty in data modeled with probability distributions. Although Monte Carlo (MC) sampling techniques have been used in prior studies to quantify critical point uncertainty, they are often expensive and are infrequently used in production-quality visualization software. We, therefore, propose a new end-to-end framework to address these challenges that comprises a threefold contribution. First, we derive the critical point uncertainty in closed form, which is more accurate and efficient than the conventional MC sampling methods. Specifically, we provide the closed-form and semianalytical (a mix of closed-form and MC methods) solutions for parametric (e.g., uniform, Epanechnikov) and nonparametric models (e.g., histograms) with finite support. Second, we accelerate critical point probability computations using a parallel implementation with the VTK-m library, which is platform portable. Finally, we demonstrate the integration of our implementation with the ParaView software system to demonstrate near-real-time results for real datasets.
}

\keywords{Topology, uncertainty, critical points, probabilistic analysis}

\teaser{
  \centering
  \includegraphics[width=\linewidth]{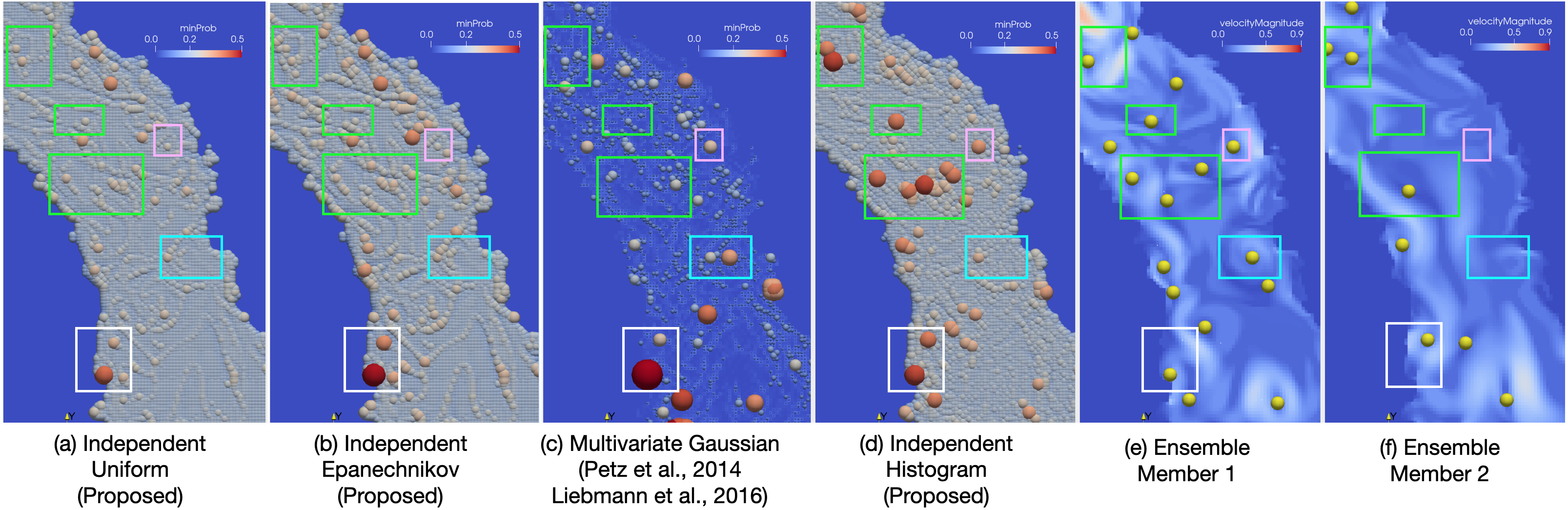}
  \vspace{-6mm}
  \caption{%
  	Parametric (a-c) vs. nonparametric (d) models for uncertainty visualization of local minima of the magnitude of velocity field vectors in the Red Sea ensemble simulations that correspond to eddy-like features. Images (e-f) show two randomly picked ensemble members with critical points rendered in yellow. The proposed nonparametric models show new high-probability features (larger spheres in red and orange) in green boxes that are not highlighted in parametric models. The results of nonparametric models can be trusted more because of the ability of nonparametric models to capture the shapes of multimodel and skewed distributions~\cite{TA:Kai:2013:nonparametricIsoVis,TA:Athawale:2016:nonparametricIsosurfaces}, unlike the restricted shape assumptions for the parametric models. The pink box shows a feature that is revealed by the multivariate Gaussian and histogram models but not by others. The cyan boxes show high-probability critical points in the multivariate Gaussian model that are not displayed as prominently in other models. The white boxes show two critical points that consistently have high probability across all models. All visualizations are computed efficiently and visualized in ParaView~\cite{TA:2005:paraview} using VTK-m~\cite{TA:2016:Moreland:vtkm} as a backend. 
  }
  \vspace{-1mm}
  \label{fig:teaser}
}

\graphicspath{{figs/}{figures/}{pictures/}{images/}{./}} 

\usepackage{tabu}                      
\usepackage{booktabs}                  
\usepackage{lipsum}                    
\usepackage{mwe}                       

\usepackage{mathptmx}                  

\begin{document}


\firstsection{Introduction}

\maketitle

Topological data analysis (TDA) is increasingly used in scientific visualizations because of its ability to concisely convey position and scale of important data features in complex datasets. The application of TDA can be found in diverse domains, including combustion science~\cite{Bremer09tvcg}, molecular dynamics~\cite{NatarajanMolecularDynamics}, and hydrodynamics~\cite{TiernyHydrodynamics}. Critical points are fundamental topological descriptors of scalar fields and form the basis of many topological visualization techniques, including persistent diagrams\cite{TA:Edelsbrunner:2010:computational_topology_intro}, contour trees~\cite{TA:Carr:2000:contourTrees, TA:Gyulassy:2005:scalarFieldSimplification}, and Morse complexes\cite{TA:Edelsbrunner:2003:MorseSmaleComplexes}. Critical points denote the domain position where the gradient of a field vanishes (see technical details in \cref{sec:backgroundAndproblemSetting}). \blfootnote{This manuscript has been authored by UT-Battelle, LLC under Contract No. DE-AC05-00OR22725 with the U.S. Department of Energy. The publisher, by accepting the article for publication, acknowledges that the U.S. Government retains a non-exclusive, paid up, irrevocable, world-wide license to publish or reproduce the published form of the manuscript, or allow others to do so, for U.S. Government purposes. The DOE will provide public access to these results in accordance with the DOE Public Access Plan (\url{http://energy.gov/downloads/doe-public-access-plan}).} Uncertainty inherent in data arising from instrument/simulation/model errors~\cite{TA:Brodlie:2012:RUDV}, however, creates uncertainty regarding critical point positions. Ignoring uncertainty in critical point positions, therefore, can lead to errors in topological visualizations and analysis. Thus, it is necessary to quantify and visually convey uncertainty in critical points to prevent misinformation. In this paper, we study uncertainty in critical points as a function of uncertainty in the underlying data modeled with probability distributions. 

A significant number of studies have investigated uncertainty in critical points arising from uncertain scalar fields. Petz et al.~\cite{Petz2012criticalPointProbability} and Liebmann and Scheuermann~\cite{Liebmann2016CriticalPointUncertainty} modeled noise in data with multivariate Gaussian distribution to derive critical point probabilities. The former work utilized Monte Carlo (MC) sampling of the original space, and the latter work utilized MC sampling of informative subspaces (referred to as patch sampling in their work) for deriving critical point probabilities. G\"{u}nther et al.~\cite{DavidJosephJulien2014} investigated spatial bounds in the domain where at least one critical point is guaranteed to exist for the uniform noise assumption. Mihai and Westermann~\cite{MIHAI201413}
derived confidence intervals for gradient field and Hessian to visualize likely critical point positions and their type in the domain. Vietinghoff et al.~\cite{VietinghoffBaysianCriticalPointUncertainty} derived the critical point probability using the Bayesian inference and derived confidence intervals~\cite{VietinghoffCriticalPointCIs}. Recently, Vietinghoff et al. developed a novel mathematical framework~\cite{VietinghoffMathematicalFoundationCriticalPoints} that quantified uncertainty in critical points by analyzing the variation in manifestation of the same critical points occurring across realizations of the ensemble.

Inspired by these advances, we propose a novel closed-form theoretical framework for uncertainty quantification and visualization of critical point uncertainty. In particular, we analytically derive the critical point probability per grid position in a regular grid for independent parametric and nonparametric noise models with finite support. Although the previously proposed multivariate Gaussian noise models~\cite{Petz2012criticalPointProbability, Liebmann2016CriticalPointUncertainty} can better handle the noise correlation compared to independent noise models, they have two shortcomings. First, they resort to the MC sampling approach for deriving the critical point probability, which can be computationally expensive depending on the number of samples and grid resolution, and they converge slowly to the true answer. Second, because of the restricted shape assumptions for the (parametric) Gaussian distributions, they can be less robust to the data outliers than the nonparametric noise models~\cite{TA:Kai:2013:nonparametricIsoVis, TA:Athawale:2021:nonparametricDVR}. We address these two shortcomings by developing a closed-form formulation and algorithm that enhance the efficiency of results compared to MC sampling. Further, we showcase how the nonparametric models can provide more robust results compared to multivariate Gaussian and other parametric models. 

Another important challenge associated with uncertainty visualization algorithms is the additional computational costs they bear, which prevents their integration with production-level software, e.g, VisIt~\cite{VisIt} and ParaView~\cite{TA:2005:paraview}. The problem of the added cost is amplified for expensive MC sampling methods (e.g., in the case of multivariate Gaussian models). It is, therefore, important to research techniques that facilitate the integration of uncertainty visualization with production-level tools to make them usable and accessible to a broader scientific community. Recently, Wang et al.~\cite{Wang2023} provided a parallel and platform-portable implementation of isosurface uncertainty visualization using the VTK-m library~\cite{TA:2016:Moreland:vtkm}. They also showcased the integration of their VTK-m implementation with ParaView. The Topology Toolkit (TTK) by Tierny et al.~\cite{TA:2018:TFL} provided efficient implementation of mandatory critical points~\cite{DavidJosephJulien2014} that is usable in ParaView. Motivated by these works, we present a VTK-m parallel implementation of our closed-form critical point probability computations that is integrable with ParaView. 

To summarize, our contributions are threefold. First, we propose a theoretical framework for uncertainty computation of critical points in uncertain 2D scalar fields. In particular, we propose an algorithm for deriving local minimum, local maximum, and saddle probability in closed form when data uncertainty is modeled as independent parametric (uniform, Epanechnikov) and nonparametric (histogram) distributions with finite support. Second, we evaluate our algorithms by demonstrating their enhanced accuracy and comparing their performance with the conventional MC models. We showcase the increased robustness of our proposed nonparametric models to data outliers compared to parametric models through results on a synthetic dataset. We present the utility of our methods through experiments on real datasets. Lastly, we implement our algorithms using the VTK-m library to present accelerated computation of critical point uncertainty and demonstrate their usability in ParaView for broader community access.
\section{Related Work}\label{relatedwork}

The research in uncertainty visualization dates back to early 2000 when Pang~\cite{TA:Pang97approaches} and Johnson~\cite{TA:Johnson:2004:topSciVisProblems, TA:Johnson:2003:nextStepVisErrors} recognized the need for quantifying and depicting uncertainty in visualization. Since then, multiple advances in uncertainty visualization have been documented in multiple survey reports, including those by Brodlie et al. \cite{TA:Brodlie:2012:RUDV}, Potter et al.~\cite{TA:Potter:2012:UQtaxonomy}, and Kamal et al.~\cite{TA:Kamal:2021:UQvisSurvey}. Uncertainty visualization specific to ensemble data and topology was discussed in recent survey reports by Hazarika et al.~\cite{TA:Wang:2019:uncertaintyEnsemble} and Yan et al.~\cite{TA:Yan:2021:scalarFieldTopoComparison}, respectively. 
 
Multiple new techniques have been derived to portray uncertainty in scalar, vector, tensor, and multivariate data. Uncertainty visualization of scalar field data covers a range of algorithms, including isosurfaces of univariate data~\cite{TA:Pothkow:2011:probMarchingCubes, Wang2023, TA:Athawale:2016:nonparametricIsosurfaces, TA:2016:Ferstl:isosurfaceUncertainty, TA:2022:Han}, multivariate surfaces~\cite {TA:Athawale:2023:uncertainFibers, TA:Sane:2021:uncertainFeatureLevelSets}, direct volume rendering~\cite{Djurcilov:2002:dvrUncertaintyVis,Athawale:2013:linGaussian, TA:Shusen:2012:GMMdvr,TA:Athawale:2021:nonparametricDVR}, topological merge trees~\cite{TA:2020:Yan:mergeTreeAverage}, contour trees~\cite{TA:Wu:2013:ContourTreeUncertainty}, persistence diagrams~\cite{TA:2020:Vidal:persistenceDiagramBarycenter}, and Morse complexes~\cite{TA:Athawale2022MsComplex}. Although not to the extent of scalar fields, vector-field uncertainty has been explored to gain insight into important data features, including critical point uncertainty~\cite{Otto:2010:uncertain2Dfield,TA:Otto:2011:3dVectorFieldTopologyUncertainty}, streamlines~\cite{TA:Ferstl:2016:streamlineVariabilityVis, TA:2016:He:probabilisticStreamlines}, and Finite-Time Lyapunov Exponents~\cite{TA:2016:Guo:FTLE}. A few techniques have been developed to compute and visualize uncertainty in tensor field data captured with high angular resolution diffusion imaging (HARDI)~\cite{jiao:TensorUncertainty:2012} and diffusion tensor imaging (DTI)~\cite{dTIuncertainty:siddiqui:2021}. These prior contributions mainly included Monte Carlo sampling, Bayesian statistics, closed-form solutions, empirical ensemble analysis, and low-dimensional embedding techniques to understand uncertainty in data features. The methods proposed in this paper model uncertain 2D scalar data as a probabilistic field and derive closed-form solutions to understand critical point uncertainty.

A variety of noise models have been previously explored to visualize uncertainty in scientific data. Statistically independent Gaussian~\cite{pothkowIndependentGaussianDataIsocontours} and uniform~\cite{TA:Athawale:2013:lerpUncertainty, DavidJosephJulien2014} distributions have been used to model uncertainty in scalar fields and study their impact on features, such as level-sets and critical points. This work was later extended to multivariate Gaussian noise models~\cite{TA:Pothkow:2011:probMarchingCubes, TA:2022:Han, dataDrivenMultivariateGaussianUncertainty, fast-spatial-levelCrossingProbability, Liebmann2016CriticalPointUncertainty} to capture the correlation among uncertain data and avoid overestimation of feature probabilities in the final visualization caused by the data independence assumption. Nonparametric models (e.g., histograms, kernel density estimation, Gaussian mixture models) have been used to show enhancements in visualization quality over parametric models for various visualization techniques, including level-sets~\cite{TA:Kai:2013:nonparametricIsoVis,TA:Athawale:2016:nonparametricIsosurfaces}, direct volume rendering~\cite{TA:Shusen:2012:GMMdvr, TA:Athawale:2021:nonparametricDVR}, and fiber surfaces~\cite{TA:Athawale:2023:uncertainFibers}, because of their higher robustness to outliers. Nonparmaetric models, however, possess extra computational cost compared to parametric models. Recently, copula-based models have been explored to capture the correlation between both parametric and nonparametric models for uncertainty visualization~\cite{copula-uncertaintyVisualization}. In this paper, we present our methods by modeling data uncertainty with independent parametric and nonparametric noise models with finite support.

Effective presentation of uncertainty is another important research challenge. Visual attribute-mapping techniques, e.g.,  colormapping~\cite{Rhodes:2003:isosurfaceUncertaintyVis} and point movement~\cite{TA:Grigoryan:2002:probabilisticSurfaces} proportional to uncertainty, have been proposed to convey uncertainty in 3D surfaces. We utilize the elevation map technique proposed by Petz et al.~\cite{Petz2012criticalPointProbability} to render critical point uncertainty for the climate dataset in \cref{sec:results}. The novel use of glyphs has been previously proposed for conveying uncertainty in vector~\cite{glyphVectorUncertainty:1996:Wittenbrink} and tensor~\cite{TA:Jones:2003:uncertaintyConeGlyphsTractography} field data. We utilize sphere glyphs to show critical point uncertainty for the Red Sea dataset in \cref{fig:teaser}. Animation techniques have been proposed for volume rendering~\cite{animationUncertainty:2007}. Effective quantification and visualization of uncertainty for 2D and high-dimensional data still remains a big challenge for the visualization community.

\section{Background and Problem Setting} \label{sec:backgroundAndproblemSetting}

 We briefly define critical points. Let $\Mspace \subset \Rspace^2$ be a 2D domain with a boundary discretized as a regular grid  (we further ignore the boundary condition for most of our discussion). Let $f: \Mspace \to \Rspace$ be a function; $\grad{f}$ denotes its gradient. A point $p \in \Mspace$ is considered {\em critical} if $\grad{f}(p)=0$; otherwise it is {\em regular}. We will assume that all critical points are non-degenerate, i.e., $f$ is a Morse function. A critical point is categorized into three types, local minimum ($l_{min}$), local maximum ($l_{max}$), and saddle ($l_s$). In particular, if $f(p)$ is smaller than the function value of all of its neighbors, then the point $p$ is a local minimum. Similarly, if $f(p)$ is greater than the function value of all of its neighbors, then the point $p$ is a local maximum. If $f(p)$ is smaller than the function value of one neighbor and greater than the one for the next neighbor in alternating fashion with the neighbors visited sequentially in clockwise or counterclockwise manner, then the point $p$ is a saddle.    

\begin{figure}[!ht]
  \centering 
  \vspace{-3mm}
  \includegraphics[width=0.68\columnwidth]{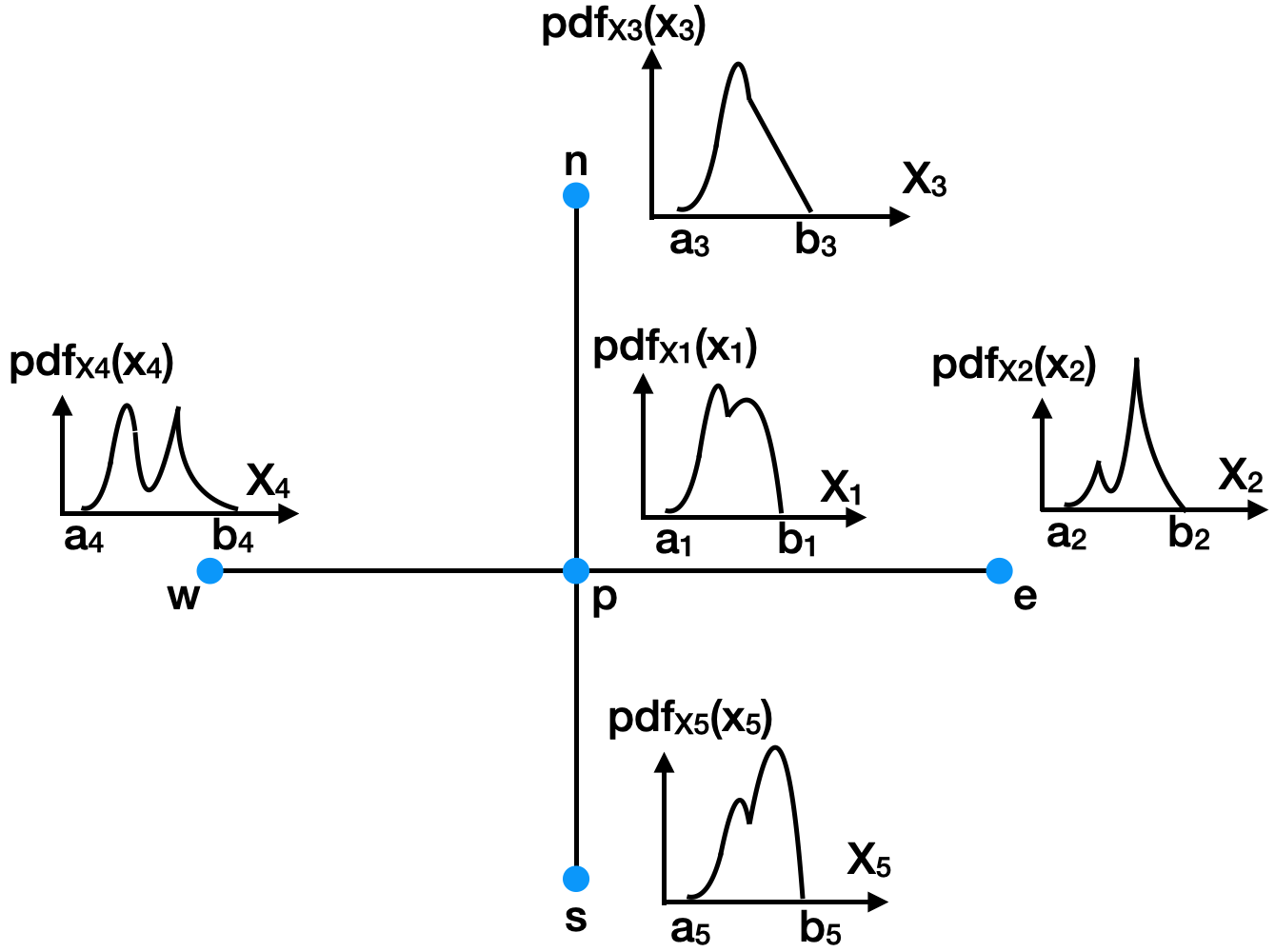}
  \vspace{-2mm}
  \caption{Depiction of a problem setting. The probability distributions at a grid vertex $p$ and its neighbors $e, n, w$, and $s$ represent uncertainty in data. Our aim is to compute the probability of point $p$ to be critical when distributions are represented with parametric and nonparametic models.
  }
  \vspace{-2mm}
  \label{fig:problemSetting}
\end{figure}
 Uncertainty in data, however, creates uncertainty regarding whether a point is critical or not. In this paper, we calculate the probability for a domain point to be critical when uncertainty in data is modeled as probability distributions. Specifically, the methods in this paper consider uncertain data at four neighbors of a domain point along the two coordinate axes directions in a regular grid to compute the probability of the domain point to be critical. Although several applications consider six- or eight-pixel neighborhoods depending on the domain triangulation for critical point visualization, we plan to research these cases with a higher number of neighbors in the future. 
 
 \cref{fig:problemSetting} depicts the problem setting, which is also used to introduce notation. Let $p$ be a point for which we want to compute the probability of it being critical. Let $X_1 \sim \text{Pdf}_{X_1}(x_1)$ denote a random variable with parametric or nonparametric noise distribution $\text{Pdf}_{X_1}$ over the support $x_1 \in [a_1,b_1]$ at a point $p$. Let $e, n, w$, and $s$ be the four neighbors of a point $p$ in the east, north, west, and south directions with random variables $X_2 \sim \text{Pdf}_{X_2}(x_2)$, $X_3 \sim \text{Pdf}_{X_3}(x_3)$, $X_4 \sim \text{Pdf}_{X_4}(x_4)$, and $X_5\sim \text{Pdf}_{X_5}(x_5)$, respectively, that denote uncertainty in data. For each random variable $X_i$, $x_i \in [a_i,b_i]$ with $a_i < b_i$. Our work presents all derivations for the independent noise models with noise distribution over a finite support, i.e., the random variables $X_i$ with $i \in \{1, \dots, 5\}$ are assumed to be independent and bounded by a finite support $[a_i,b_i]$. As the local data are not always independent in real datasets and identifying bounds $[a_i,b_i]$ can be challenging considering noisy data acquisition processes~\cite{TA:Brodlie:2012:RUDV}, we discuss the ramifications of our independent noise and finite support assumptions in \cref{sec:conclusion}. Because of the data independence assumption, the joint probability density $\text{Pr}_{joint}$ of the random variables is the product of their individual probability densities, i.e., $\text{Pdf}_{joint} = \prod_{i=1}^{i=5}\text{Pdf}_{X_i}(x_i)$. Let $\diff x = \prod_{i=1}^{i=5} \diff x_i$. Given these data, our goal is to find the probability of point $p$ being a local minimum $\text{Pr}(p=l_{min})$, local maximum $\text{Pr}(p=l_{max})$, and saddle $\text{Pr}(p=l_s)$. 
\section{Methods}
We describe the mathematical formulation and our algorithm for critical point probability computation in closed form for independent  parametric and nonparametric noise models with finite support.

\subsection{Critical Point Probability (Two-Pixel Neighborhood)}\label{sec:criticalPointProbability-TwoNeighborhood}
For simplicity, we describe our derivations and approach for computing the probability of point $p$ to be critical for 1D uncertain scalar fields. Our methods for the 1D case generalize to uncertain 2D scalar fields, as described in \cref{sec:criticalPointProbability-FourNeighborhood}. For the 1D case, we consider only the two neighbors with random variables $X_2$ and $X_3$ of a 1D point $p$ with its associated random variable $X_1$. The rest of the problem settings are similar to the 2D case described earlier in \cref{sec:backgroundAndproblemSetting}. 

\subsubsection{Local Minimum Probability}\label{sec:1D-localMinProb}

The probability of point $p$ being a local minimum, $\text{Pr}(p=l_{min})$, can be computed by integrating the joint probability $\text{Pr}_{joint}$ over its support where random variable $X_1$ is simultaneously smaller than all neighboring random variables (i.e., $X_2$ and $X_3$). Mathematically, the local minimum probability can be represented as follows:
\begin{equation}\label{eq:1D-localMinimumProb}
\begin{split}
\text{Pr}(p=l_{min}) &= \text{Pr}[(X_1 < X_2) \text{ and } (X_1<X_3)]\\
& = \int_{x_1=a_1}^{x_1=b_{min}}\int_{x_2=max(x_1,a_2)}^{x_2=b_2}\int_{x_3=max(x_1,a_3)}^{x_3=b_3} (\text{Pdf}_{joint}) \diff x \text{,}\\
& \text{where } b_{min} = min(b_1,b_2,b_3)
\end{split}
\end{equation}
\noindent
\Cref{eq:1D-localMinimumProb} represents the core integration formula for the computation of local minimum probability at a domain position $p$. We describe our approach for the computation of local minimum probability in three parts. (1)~We explain the integral limits and piecewise simplifications of the core integration formula in \cref{eq:1D-localMinimumProb}. (2)~We describe our piecewise integration approach to efficiently compute the formula in \cref{eq:1D-localMinimumProb}. (3)~We show a running illustration of our piecewise integration approach to compute the local minimum probability.

\paragraph{Limits \scalebox{1.1}{$a_1$} and \scalebox{1.1}{$b_{min}$} of the outer integral in~\cref{eq:1D-localMinimumProb}:} The outer integral of~\cref{eq:1D-localMinimumProb} indicates the portion of data range of a random variable $X_1$ (i.e., $[a_1, b_1]$) that can result in point $p$ being a local minimum. In particular, the portion $[a_1, b_{min}]$ (with $a_1 < b_{min}$) of random variable $X_1$ can result in point $p$ being a local minimum, where $b_{min}$ denotes the minimum among $b_1, b_2$, and $b_3$. In contrast, the data range $[b_{min}, b_1]$ for $b_{min} \neq b_1$ cannot result in a point $p$ as a local minimum $(l_{min})$ because it will be always greater than the random variables $X_2$ or $X_3$ depending on if $b_{min}=b_2$ or $b_{min}=b_3$, respectively. Thus, mathematically, for any value $x_1 \ge b_{min}$, $\text{Pr}(p=l_{min}) = 0$. In the case $b_{min} < a_1$, then $Pr(p=l_{min})=0$ because there is at least one random variable between $X_2$ and $X_3$ that will be always smaller than $X_1$.

\paragraph{Limits \scalebox{1.1}{$max(x_1,a_i)$} and \scalebox{1.1}{$b_i$} of the inner integrals in~\cref{eq:1D-localMinimumProb}:} The two inner integrals in \cref{eq:1D-localMinimumProb} integrate the joint distribution $\text{Pdf}_{joint}$ over its support where random variables $X_2$ and $X_3$ are simultaneously greater than $x_1 \in [a_1, b_{min}]$ in the outer integral. The inner integral lower limits are $max(x_1,a_i)$ for $i \in \{2,3\}$. The maximum is taken because the support of a random variable $X_i$ is restricted to $[a_i,b_i]$. Thus, for $x_1 < a_i$ in any inner integral, the entire support $[a_i,b_i]$ with $i \in \{2,3\}$ will always be greater than $x_1$, and the inner integral does not depend on the value of $x_1$. In contrast, for $x_1 > a_i$, the inner integration depends on the value of $x_1$ because $x_1$ assumes values in support of distributions. It is guaranteed that the upper limit of inner integrals in \cref{eq:1D-localMinimumProb} is greater than their respective lower limit, i.e., $b_i \ge max(x_1,a_i)$, for two reasons. First, for any random variable $X_i$, we assume $a_i < b_i$. Second, the maximum value of $x_1$ is equal to $b_{min}=min(b_1,b_2,b_3)$ based on the outer integral (see the previous paragraph), and it cannot exceed the upper limits $b_2$ and $b_3$ in inner integrals. Depending upon whether the $max(x_1,a_i)$ is equal to $x_1$ or $a_i$, the integral in \cref{eq:1D-localMinimumProb} can be simplified and computed differently, which necessitates the evaluation of the integral in \cref{eq:1D-localMinimumProb} in a piecewise manner, as described next. 

\begin{figure*}[!ht]
  \centering 
  \includegraphics[width=\linewidth]{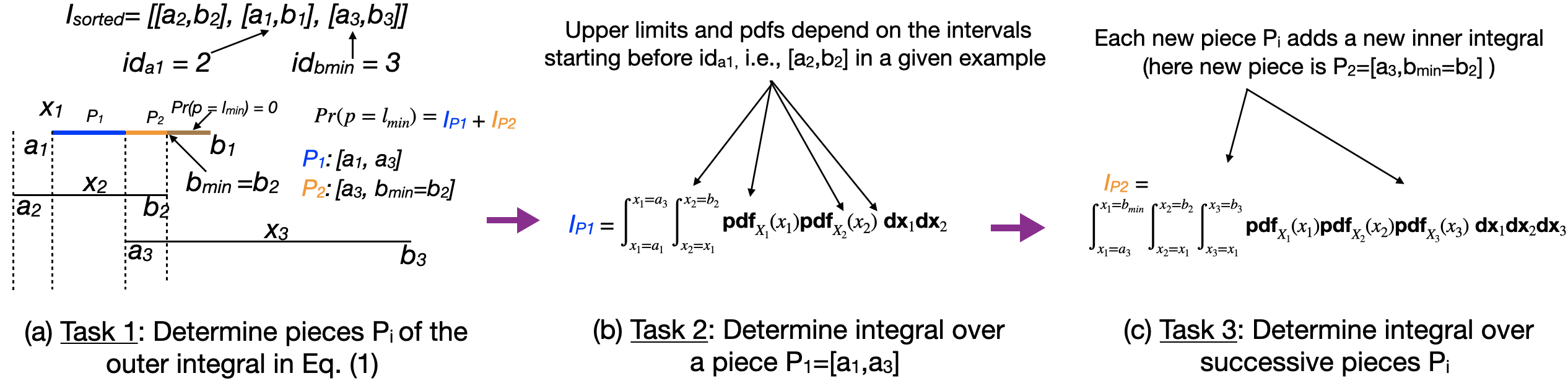}
    \vspace{-5.5mm}
  \caption{Illustration of our three-step approach for critical point uncertainty computation for the 1D case. (1) Determine the range of random variable $X_1$ for which critical point probability is nonzero (i.e., $[a_1, b_{min} = b_2]$) and determine its pieces. The range for which critical point probability is zero is shown in brown in (a). Each new start point $a_i \in [a_1, b_{min}]$ creates a new piece. Here, $a_3$ results in two pieces $P_1 = [a_1,a_3]$ and $P_2 = [a_3,b_{min}]$. (2) Compute integration for piece $P_1$ (i.e., $I_{P_1}$) depending on which intervals overlap it. Since the interval $[a_2,b_2]$ overlaps with $P_1$, $I_{P_1}$ corresponds to integral over a joint distribution of $X_1$ and $X_2$, as depicted in (b). (3) Update integrals for next pieces depending on observed start points (e.g., inclusion of random variable $X_3$ in the integral $I_{P_2}$ in (c) based on the start point $a_3$) and sum all piecewise integrals to compute the local minimum probability.
  }
  \vspace{-5mm}
  \label{fig:1D-localMinimumProbAlgorithm}
\end{figure*}
\paragraph{Piecewise simplification of~\cref{eq:1D-localMinimumProb}:} The core integration formula in \cref{eq:1D-localMinimumProb} can be simplified differently for different subsets of the range of the outer integral (i.e., $[a_1, b_{min}]$). We refer to each subset as a piece $P$. For a piece $P$, where $x_1 < a_2$ and $x_1 < a_3$, the inner integrals in~\cref{eq:1D-localMinimumProb} attain the range $x_2 \in [a_2, b_2]$ and $x_3 \in [a_3, b_3]$. In other words, the inner integrals do not depend on $x_1$ when $x_1 < a_2$ and $x_1 < a_3$. Thus, the integration over the entire support of random variables $X_2$ and $X_3$ simplifies \cref{eq:1D-localMinimumProb} to the integration over a marginal distribution of $X_1$ for the piece $P$, i.e., $\int_{x_1 \in P}\text{Pdf}_{X_1}(x_1) \diff x_1$.

For a piece $P$, where $x_1 \in [a_2,b_2]$  and $x_1 < a_3$, the inner integrals in~\cref{eq:1D-localMinimumProb} attain the range $x_2 \in [x_1, b_2]$ and $x_3 \in [a_3, b_3]$. In this case, only the first inner integral related to the range of random variable $X_2$ depends upon $x_1$. Thus, \cref{eq:1D-localMinimumProb} simplifies as the integration over the joint distribution of $X_1$ and $X_2$ for the piece $P$, i.e., $\int_{x_1 \in P}\text{Pdf}_{X_1}(x_1) \text{Pdf}_{X_2}(x_2)\diff x_2\diff x_1$. In summary, various pieces of the outer integral in~\cref{eq:1D-localMinimumProb} can be simplified differently based on whether the inner integrals depend on $x_1$ or not.

\paragraph{Approach for computing local minimum probability:} The computation of \cref{eq:1D-localMinimumProb} depends on the ordering of start points $a_i$ (i.e., $a_1, a_2$, and $a_3$) and $b_{min}$. Thus, in 1D case, there are $4!=24$ permutations of $a_i$ and $b_{min}$. These permutations increase fast for 2D/high-dimensional cases. We, therefore, devise an efficient algorithm that computes the piecewise integrals on the fly depending on the observed permutation of $a_i$ and $b_{min}$ without needing to go through all permutations.  

We now describe our approach for the computation of the local minimum probability at a domain position $p$ [i.e., $\text{Pr}(p = l_{min})$]. Our approach comprises three main tasks. \textbf{Task 1: Determination of pieces $P_i$ of the outer integral in \cref{eq:1D-localMinimumProb} needed for performing piecewise integration.} Initially, we compute $b_{min}$. If $b_{min} < a_1$, then there are no pieces and $\text{Pr}(p = l_{min})=0$. If $b_{min} > a_1$, then we sort intervals representing uncertain data ranges (i.e., $[a_i,b_i]$) based on their start points $a_i$ and keep them in the array named $I_{sorted}$. We note the index of the interval $[a_1,b_1]$ in $I_{sorted}$ (referred to as $id_{a_1}$) and the index of interval in $I_{sorted}$ from the end that contains $b_{min}$ (referred to as $id_{b_{min}}$), as $a_1$ and $b_{min}$ constitute limits of the outer integral in \cref{eq:1D-localMinimumProb}. Any start points $a_i$ contained in the range of indices $id_{a_1}$ and $id_{b_{min}}$ determine the pieces $P_i$ for integration.  This process generalizes to any ordering of $a_i$ to determine the pieces $p_i$ of the outer integral in \cref{eq:1D-localMinimumProb}. 

Next, we compute the integration for piece $P_1$ denoted as $I_{P_1}$. \textbf{Task 2: Integration over piece $P_1$ of the outer integral range \bm{$[a_1, b_{min}]$}.} The integration for piece $P_1$ depends on the intervals that started before $a_1$ because the inner integral in \cref{eq:1D-localMinimumProb} depends on $x_1$ for a random variable $X_i$ (with $i \in {2,3}$) started before $a_1$, as max($x_1$,$a_i$) is equal to $x_1$. Task 2, therefore, corresponds to finding the intervals that started before $id_{a_1}$, which also determines the upper limits of inner integrals for piece $P_1$ depending on observed order of intervals. 

The computation of the integral in \cref{eq:1D-localMinimumProb} for piece $P_1$ (as well as any arbitrary piece $P_i$) simplifies to one of the three types of integration formulae, which we call integration templates. Generally, the simplification depends on the number of intervals overlapping a piece $P_i$, as explained earlier. If there is no overlap with a piece, then the integral in \cref{eq:1D-localMinimumProb} simplifies to the integration of the probability distribution of random variable $X_1$ (Template 1) over a piece. If only one random variable ($X_2$ or $X_3$) is overlapping with a piece, then the integral in \cref{eq:1D-localMinimumProb} simplifies to the integral over the joint distribution of $X_1$ and a random variable corresponding to the overlapping interval (Template 2). If both random variables  $X_2$ and $X_3$ are overlapping with a piece, then the integral in \cref{eq:1D-localMinimumProb} corresponds to the integral over the joint distribution of all random variables (Template 3).

Having determined the integration for piece $P_1$, we compute integration for successive pieces. \textbf{Task 3: Integration over piece $P_i$ with $i>1$.} Essentially, each new start point $a_i$ observed between the outer integral limits $a_1$ and $b_{min}$ of \cref{eq:1D-localMinimumProb} creates a new piece. Generally speaking, each new start point $a_i$ results in a different simplification of \cref{eq:1D-localMinimumProb} because $max(x_1, a_i)$ in \cref{eq:1D-localMinimumProb} becomes equal to $x_1$ at each new start point. Thus, encountering a new start point adds one inner integral in a simplified form compared to the piece before encountering a new start point. Finally, the integration of all pieces is summed to compute the local minimum probability at a point $p$, i.e., $\text{Pr}(p=l_{min})$. 

\paragraph{Illustration of Local Minimum Probability Computation:}

\cref{fig:1D-localMinimumProbAlgorithm} illustrates our method for computing
the local minimum probability for a domain point $p$. As shown for the example in \cref{fig:1D-localMinimumProbAlgorithm}, $a_2 < a_1 < a_3 < b_2 < b_1 < b_3$. Initially, we determine the range of a random variable $X_1$ that can result in point $p$ being a local minimum. As shown in \cref{fig:1D-localMinimumProbAlgorithm}a, each value in the range $[x_1=a_1,x_1=(b_{min}=b_2)]$ has a nonzero probability of being simultaneously smaller than the neighboring random variables (i.e., $X_2$ and $X_3$). On the contrary, the range $[x_1=(b_{min}=b_2), x_1=b_1]$ is always greater than the random variable $X_2$, and therefore, cannot result in point $p$ as a local minimum.

In Task 1, we determine the pieces $P_i$ of the range $[x_1=a_1,x_1=(b_{min}=b_2)]$. As depicted in \cref{fig:1D-localMinimumProbAlgorithm}, the array $I_{sorted}$ has intervals ordered by $a_i$, where $a_2 < a_1 < a_3$. For this $I_{sorted}$, $id_{a_1}=2$ and $id_{b_{min}}=3$. Since $a_3$ is a start point in interval indexed by $id_{b_{min}}$, it divides the outer integral range $[a_1,b_{min}]$ in \cref{eq:1D-localMinimumProb} into two pieces (depicted in blue and orange in \cref{fig:1D-localMinimumProbAlgorithm}).

In Task 2, we determine the simplification of the formula in \cref{eq:1D-localMinimumProb} for piece $P_1$. The simplification for piece $P_1 = [a_1,a_3]$ in \cref{fig:1D-localMinimumProbAlgorithm} (denoted by blue) depends on the number of intervals that started before $a_1$. 
As observed in \cref{fig:1D-localMinimumProbAlgorithm}, the interval $[a_2,b_2]$ starts before $a_1$.  Since $x_1 < a_3$ in piece $P_1$, the formula in \cref{eq:1D-localMinimumProb} integrates random variable $X_3$ over its entire support and simplifies to the integration over the joint distribution of random variables $X_1$ and $X_2$ shown in \cref{fig:1D-localMinimumProbAlgorithm}b.

In Task 3, we determine the simplification of the formula in \cref{eq:1D-localMinimumProb} for successive pieces formed by each new start point $a_i \in [a_1, b_{min}]$. In \cref{fig:1D-localMinimumProbAlgorithm}, the start point $a_3 \in [a_1,b_{min}]$ results in a new piece $P_2$ shown in orange.  All inner integrals for piece $P_2$ stay the same as piece $P_1$ except for one newly added inner integral with limits $[x_1,b_3]$, as shown in \cref{fig:1D-localMinimumProbAlgorithm}c, because $x_1 = max(x_1,a_3)$ for piece $P_2$, unlike the piece $P_1$ in which $a_3 = max(x_1,a_3)$. Thus, we make such updates to inner integrals for each new piece corresponding to a new start point $a_i \in [a_1, b_{min}]$.

\paragraph{Time complexity:}Our presented approach for local minimum probability computation initially sorts all the intervals based on start points $a_i$ with $i \in {1, 2, 3}$ and $b_{min}$ to determine pieces for integration (Task 1), which is a constant time operation. Task 2 and Task 3 comprise a single loop, which runs a maximum of three times corresponding to the three entries of a sorted interval array $I_{sorted}$. Each loop iteration computes the integral template (simplification of the formula in \cref{eq:1D-localMinimumProb}) on the fly depending on the observed data in array $I_{sorted}$ in constant time. The algorithm is, therefore, linear time complexity with the number of input intervals (here, three) and extremely efficient. 
\subsubsection{Local Maximum Probability}\label{sec:1D-localMaxProb}
Having derived a probabilistic framework for computation of local minimum probability (\cref{sec:1D-localMinProb}), the computation of the local maximum probability $Pr(p=l_{max})$ at a domain point $p$ is fairly straightforward. Computation of the local maximum probability corresponds to computing $\text{Pr}[(X_1 > X_2) \text{ and } (X_1>X_3)]$ for the 1D case, which is equivalent to computing $\text{Pr}[(-X_1 < -X_2) \text{ and } (-X_1 < -X_3)]$. This negation format is equivalent to \cref{eq:1D-localMinimumProb}. Thus, we create negated random variables $X_1' = -X_1$, $X_2' = -X_2$, and $X_3' = -X_3$. We then apply our proposed local minimum probability computation algorithm (\cref{sec:1D-localMinProb}) to these new random variables $X_i'$ for computing the local maximum probability.
\subsubsection{Saddle Probability}\label{sec:1D-saddleProb}
The probability of point $p$ being a saddle, $\text{Pr}(p=l_{s})$, can be computed by integrating the joint probability $\text{Pdf}_{joint}$ over its support where the random variable $X_1$ is simultaneously smaller than $X_2$ and greater than $X_3$ (and the other way around). Mathematically, the saddle probability can be represented as follows:
\setlength{\abovedisplayskip}{8pt}
\setlength{\belowdisplayskip}{8pt}
\begin{align*}
\begin{split}
\text{Pr}(p=l_{s}) =& (t_1 = \text{Pr}[(X_1 < X_2) \text{ and } (X_1 > X_3)])+ \\
 &(t_2 = \text{Pr}[(X_1 > X_2) \text{ and } (X_1 < X_3)])\\
\end{split}
\end{align*}
\noindent
The term $t_1$ in the equation above can be written as follows: 
\begin{equation}\label{eq:1D-saddleProb}
\begin{split}
& t_1 = \text{Pr}[(X_1 < X_2) \text{ and } (X_1>X_3)]\\
& = \int_{x_1=a_{max}^{alt}}^{x_1=b_{min}^{alt}}\int_{x_2=max(x_1,a_2)}^{x_2=b_2} \int_{x_3=a_3}^{x_3=min(x_1,b_3)}(\text{Pdf}_{joint}) \diff x \text{,}\\
&\text{where } a_{max}^{alt}=max(a_1, a_3),b_{min}^{alt} = min(b_1,b_2)
\end{split}
\end{equation}
\noindent
\Cref{eq:1D-saddleProb} represents the core integration formula for the computation of the saddle probability at a domain point $p$. We derive our closed-form computations and algorithm only for the term  $t_1 = \text{Pr}[(X_1 < X_2) \text{ and } (X_1 > X_3)]$. The term $t_2$ can be computed by creating negated random variables $X_i'=-X_i$ with $i \in \{1,2,3\}$ and plugging $X_i'$ in place of $X_i$ into the derivation for term $t_1$. 
 
The algorithm to compute the saddle probability is similar to the three tasks of the local minimum probability computation in \cref{sec:1D-localMinProb}. In Task 1, the algorithm first determines the data range $[a_{max}^{alt}, b_{min}^{alt}]$ that qualifies for point $p$ being a saddle (similar to the range determination for local minimum probability computation, illustrated in \cref{fig:1D-localMinimumProbAlgorithm}a). The algorithm then divides this range into pieces $P_i$ depending on the ordering of points $[a_{max}^{alt}, b_{min}^{alt}, a_2, b_3]$ that correspond to the limits of integrals in \cref{eq:1D-saddleProb}. In Task 2, the integral of piece $P_1$ is computed depending on how $a_2$ and $b_3$ are ordered with respect to the lower limit $a_{max}^{alt}$. In Task 3, the integrals of successive pieces are computed depending on the order in which $a_2$ and $b_3$ appear until the upper limit $b_{min}^{alt}$. All integrals fit into one of the templates that represent simplification of the core integration in \cref{eq:1D-saddleProb}. All piecewise integrals are finally added to compute the term $t_1$.

\subsection{Critical Point Probability (Four-Pixel Neighborhood)} \label{sec:criticalPointProbability-FourNeighborhood}
Our methods for the two-pixel neighborhood case (\cref{sec:criticalPointProbability-TwoNeighborhood}) generalize to the four-pixel neighborhood case (depicted in \cref{fig:problemSetting}) straightforwardly.  Here, we briefly discuss a few new specifics related to the four-pixel neighborhood case. The detailed algorithms and illustrations are provided in the supplementary material.
\paragraph{Local Minimum Probability:}

The core integration formula for the computation of local minimum probability at a domain position $p$ in the case of a four-pixel neighborhood takes a form similar to \cref{eq:1D-localMinimumProb}. 
\begin{equation}\label{eq:localMinimumProb}
\begin{split}
& \text{Pr}(p=l_{min}) \\
& = \text{Pr}[(X_1 < X_2) \text{ and } (X_1<X_3)\text{ and } (X_1<X_4)\text{ and } (X_1<X_5)]\\
& = \int_{x_1=a_1}^{x_1=b_{min}}\int_{x_2=max(x_1,a_2)}^{x_2=b_2}\dots\int_{x_5=max(x_1,a_5)}^{x_5=b_5} (\text{Pdf}_{joint}) \diff x \text{,}\\
& \text{where } b_{min} = min(b_1,b_2,b_3,b_4,b_5)
\end{split}
\end{equation}
\noindent
Computation of the integral in \cref{eq:localMinimumProb} has a workflow similar to the two-pixel neighborhood case. Initially, pieces of the range $[a_1, b_{min} = min(b_1 \dots b_5)]$ are determined based on the ordering of integral limits, i.e., $a_1, a_2, a_3, a_4, a_5$, and ${b_{min}}$. Based on these six points of interest, there are $6!=720$ possible permutations. We compute piecewise simplification of \cref{eq:localMinimumProb}  (templates) on the fly depending on the observed ordering of the six points without needing to go through all
permutations. Finally, all piecewise integrals are summed to compute the local minimum probability.

\paragraph{Local Maximum Probability:}
Similar to the two-pixel neighborhood case, the local maximum probability in the four-pixel neighborhood case is computed by negating the random variables followed by the application of our algorithm for computing the local minimum probability.

\paragraph{Saddle Probability}
The core integration formula for computing probability of point $p$ being a saddle, $\text{Pr}(p=l_{s})$ in the case of four-pixel neighborhood is mathematically represented as follows:
\setlength{\abovedisplayskip}{8pt}
\setlength{\belowdisplayskip}{8pt}
\begin{align*}
\begin{split}
& \text{Pr}(p=l_{s}) \\
& = (t_1 = \text{Pr}[(X_1 < X_2) \text{ and } (X_1 > X_3)\text{ and } (X_1<X_4)\text{ and } (X_1 > X_5)])+\\
& (t_2 = \text{Pr}[(X_1 > X_2) \text{ and } (X_1 < X_3)\text{ and } (X_1 > X_4)\text{ and } (X_1 < X_5)])\\
\end{split}
\end{align*}
\noindent
The term $t_1$ in the equation above takes a form similar to the two-pixel neighborhood case in \cref{eq:1D-saddleProb} and can be written as follows: 
\begin{equation}\label{eq:saddleProb}
\begin{split}
& t_1 = \text{Pr}[(X_1 < X_2) \text{ and } (X_1>X_3)\text{ and } (X_1<X_4)\text{ and } (X_1>X_5)]\\
& = \int_{x_1=a_{max}^{alt}}^{x_1=b_{min}^{alt}}\int_{x_2=max(x_1,a_2)}^{x_2=b_2} \int_{x_3=a_3}^{x_3=min(x_1,b_3)}\int_{x_4=max(x_1,a_4)}^{x_4=b_4} \int_{x_5=a_5}^{x_5=min(x_1,b_5)} \dots  \\ 
&(\text{Pdf}_{joint}) \diff x \text{, where } a_{max}^{alt}=max(a_1, a_3, a_5),b_{min}^{alt} = min(b_1,b_2,b_4)
\end{split}
\end{equation}
\noindent

Again, the saddle probability can be efficiently computed using piecewise integration with piece limits determined by ordering of the integral limits $a_{max}^{alt}$, $b_{min}^{alt}$, $a_2$, $a_4$, $b_3$, $b_5$ in \cref{eq:saddleProb}. We compute piecewise simplification of \cref{eq:saddleProb} (templates) on the fly depending on the observed ordering of the six points and sum them up to compute the saddle probability, similar to the two-pixel neighborhood case.
\subsection{Parametric Noise Models}\label{sec:parametricModels}
\begin{figure}[!b]
  \centering 
  \vspace{-4mm}
  \includegraphics[width=0.8\linewidth]{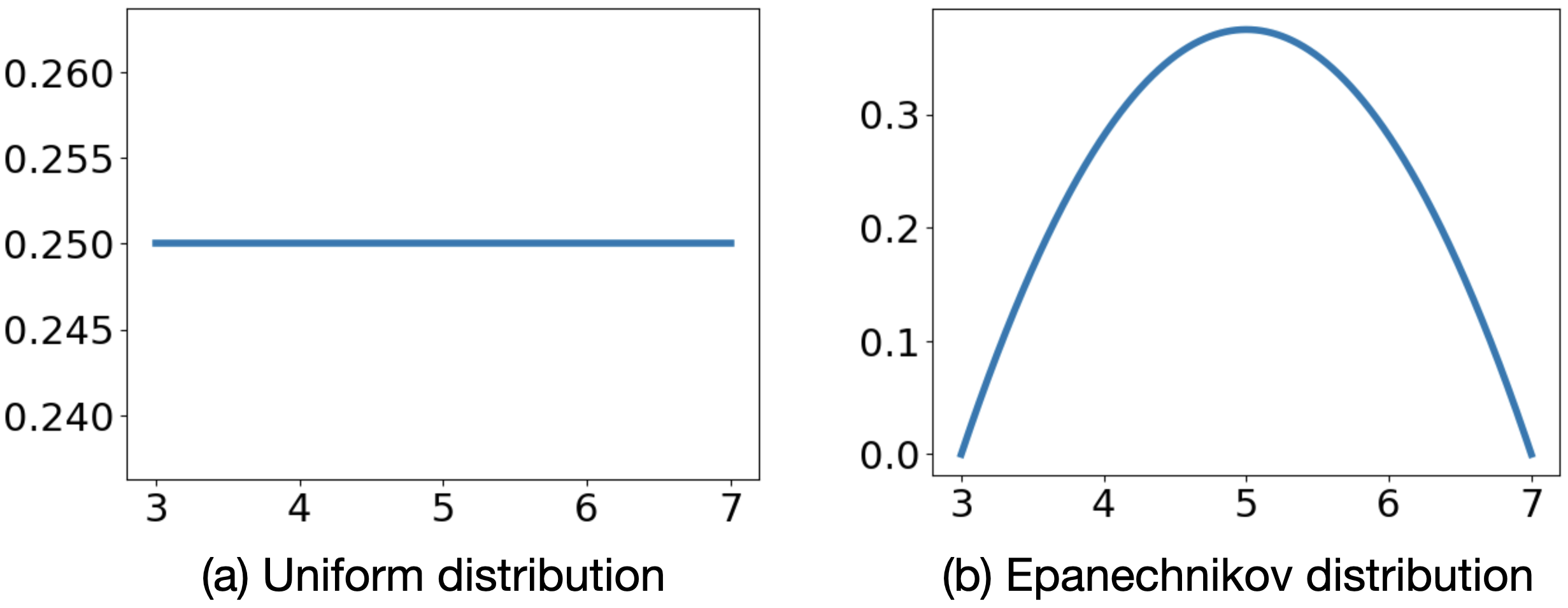}
  \vspace{-2mm}
  \caption{Epanechnikov distribution (b) gives more weight to the mean unlike the uniform distribution (a), and hence, can provide enhanced visualization compared to the uniform noise model.
  }  \label{fig:uniformVsEpanechnikov}
\end{figure}
In this section, we derive critical point probability computations for the uniform and Epanechnikov distributions. The derivation for the uniform noise model acts as a building block for our histogram model derivation in \cref{sec:closedformHistogram}. The uniform distribution is given by $\text{Pdf}_{X}(x) = \frac{1}{b-a}$, where $x \in [a,b]$. Although we do not have a closed-form solution for the Gaussian noise model, similar to the Gaussian, the Epanechnikov model gives more weight to the data mean, has a bell-like shape, and is smoother than the uniform noise distribution (see \cref{fig:uniformVsEpanechnikov}). The Epanechnikov distribution, therefore, can yield better results than the uniform noise model, as presented later in~\cref{sec:modelComparison}. The Epanechnikov distribution is given by $\text{Pdf}_X(x) = \frac{3}{2*(b-a)}[1 - (\frac{x-m}{w})^2]$, where $m=(a+b)/2$, $w=(b-a)/2$, and $x \in [a,b]$. These distribution formulae can be plugged into the integration formulae for critical point probability computations (i.e., \cref{eq:localMinimumProb} and \cref{eq:saddleProb}) and their simplifications (templates) to compute results for these two types of distributions. 

The derivation of integral templates for the uniform and Epanechnikov kernel can be cumbersome because of high-order functions resulting from integrals.  For example, since the Epanechnikov is an order-2 kernel, the integration templates can result in formulae of order-15 functions because of the five integrals in the case of four-pixel neighborhood without simplification (\cref{sec:criticalPointProbability-FourNeighborhood}). Thus, we use the Wolfram Alpha software~\cite{wolframAlpha} for deriving the integral templates\footnote{All integral templates and code are available at \url{https://github.com/tusharathawale/UCV/tree/exp_critical_point_noplugin}.}. Note that these high-order functions may produce numerical instabilities for large or fractional data values. Thus, the dataset range needs to be properly scaled to ensure stable computations with the Epanechnikov kernel.
\subsection{Nonparametric Noise Models}\label{sec:nonparametricModels}
For nonparametric models, we do not assume a particular distribution type for data. Instead, we derive a histogram with a user-specified number of bins at each grid vertex to derive critical point probabilities. Our derivations for the parametric models act as building blocks for derivations of nonparametric models. We propose closed-form and semianalytical (mix of MC and closed-form) solutions.

\subsubsection{Closed-Form Formulation}\label{sec:closedformHistogram}

Each histogram can be considered as a weighted combination of nonoverlapping uniform distributions. Mathematically, the $\text{pdf}_X(x)$ using a histogram can be written as $\text{pdf}_X(x)=w_i\sum_{i=1}^{i=h}K_{b}( x  -  x_{i})$, where $\sum_{i=1}^{i=h} w_i=1$, $K_b$ denotes a uniform kernel $K$ with width $b$, $x_i$ denotes the bin center, and $h$ denotes the number of histogram bins, and $x \in [x_1-b/2, x_h+b/2]$. Thus, our derivation and algorithm for the uniform distribution (\cref{sec:parametricModels}) can be leveraged for critical point probability computation with the histogram noise model. Let $K_{b_i}$, $K_{b_j}$, $K_{b_k}$, $K_{b_l}$, and $K_{b_m}$ denote the uniform kernels of histograms for random variables $X_1$, $X_2$, $X_3$, $X_4$, and $X_5$, respectively. The critical point probability can be computed by going through all combinations of the uniform kernels across five random variables and summing the critical point probability for each possible combination weighted by its probability. Mathematically, the local minimum probability at a grid point $p$ can be computed as follows:
\begin{equation} \label{eq:localMinProbHistogram}
\text{Pr}(p=l_{min}) = w \sum_{i=1}^{i=h} \dots \sum_{m=1}^{m=h} \text{Pr}(p = l_{min})_{i,j,k,l,m},
\end{equation}
\noindent
where $\text{Pr}(p = l_{min})_{i,j,k,l,m}$ denotes the probability of each kernel $K_{b_i}$ of random variable $X_1$ being simultaneously smaller than kernels $K_{b_j}, \dots, K_{b_m}$ of random variables $X_2, \dots, X_5$, respectively, weighted by the probability of choosing kernels denoted by $w=w_i w_j w_k w_l w_m$. We verify the correctness of our formulation through quantitative and qualitative evaluation in our results. 

Nonparametric models can increase the robustness of visualization to outliers~\cite{TA:Kai:2013:nonparametricIsoVis, TA:Athawale:2021:nonparametricDVR, TA:Athawale:2016:nonparametricIsosurfaces} because they do not assume any particular shape of the distribution. We show the increased robustness of visualizations under uncertainty with the proposed nonparametric models compared to parametric models in the results, \cref{sec:results}. However, the increased quality of nonparametric models comes at the cost of an increased number of computations. The time complexity of computing the local minimum probability is $O(h^5)$, as observed from \cref{eq:localMinProbHistogram}, where $h$ is the number of histogram bins. Our histograms method in \cref{eq:localMinProbHistogram} can be extended to a more general kernel density estimation~\cite{TA:Parzen:1968:ParzenWindow}, in which each noise sample is assigned a kernel. However, the time complexity can grow sharply with an increase in the sample count, and KDE can quickly become impractical for use in visualization. Thus, we restrict our nonparametric methods to histograms with a user-specified number of bins. To accelerate the performance of nonparametric methods, we propose two solutions. First, we present a more efficient semianalytical approach (\cref{sec:semianalytical}), which provides an approximate but reliable solution at a greater speed. Second, we accelerate our critical point uncertainty computation using a parallel implementation (\cref{sec:vtkmParaview}).

\begin{figure*}[!ht]
  \centering 
  \includegraphics[width=0.97\linewidth]{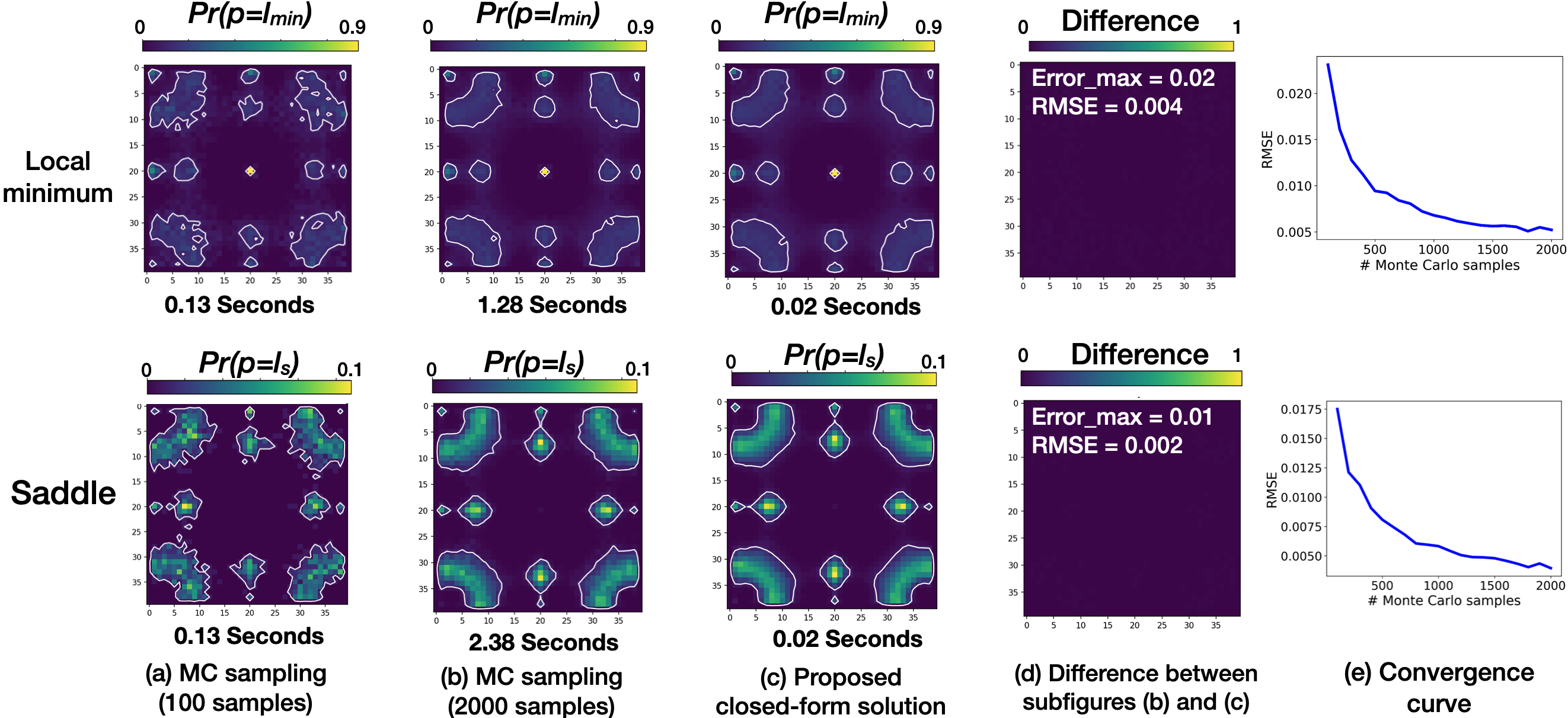}
  \vspace{-2mm}
  \caption{The qualitative and quantitative proof of correctness and enhanced performance of our proposed closed-form computations (column c) with the MC sampling approach (columns a and b) as the baseline. The results are shown for the uniform noise model. The solution obtained with 2000 MC samples converges to closed-form computations with our algorithms (see the difference images in column d and convergence curves in column e), thereby confirming their correctness. Our methods provide $64 \times$ and $119 \times$ speed-up with respect to the MC sampling approach with $2000$ samples.  
  }
  \vspace{-6mm}
  \label{fig:validationAckley}
\end{figure*}

\subsubsection{Semianalytical Solution} 
\label{sec:semianalytical}
We propose a mix of MC and closed-form formulation to get approximate but reliable solutions at a greater speed. For our method, we draw $c$ samples from $\text{pdf}_{X_1}(x_1)$ at a point $p$. For each sample, the probability of that sample being critical can be computed in closed form. Specifically, we compute the probability of a sample being smaller (or greater) than neighboring random variables in closed form. For example, in order to compute $\text{Pr}(\text{sample}<X_2)$, we first find the histogram bin of $\text{Pdf}_{X_2}(x_2)$ to which the sample belongs. We then integrate the bin density for values greater than the sample (similar to a uniform distribution) and sum it with the integration of densities of subsequent bins. Note that integration of subsequent bins is precomputed efficiently using the prefix sum (scan) method and, therefore, the integration computation time does not depend on the number of bins. Let $\text{Pr}_2 \dots \text{Pr}_5$ denote $\text{Pr}(\text{sample}<X_2) \dots \text{Pr}(\text{sample}<X_5)$, respectively. Because of the independent assumption, the local minimum probability for the sample corresponds to a product $\text{Pr}_{sample}=\text{Pr}_2\text{Pr}_3\text{Pr}_4\text{Pr}_5$. Saddle and local maximum probabilities for the sample can be computed in a similar manner. Let $\text{Pr}_{total}$ denote the sum of $\text{Pr}_{sample}$ for all $c$ samples. Thus, the probability of the point $p$ being critical can be found as the ratio of $\text{Pr}_{total}$ and  $c$. The computational complexity of this semianalytical method is proportional to the number of samples (i.e.,~$c$) drawn from a single distribution $\text{Pdf}_{X_1}(x_1)$. This method, therefore, provides faster results compared to the exponential closed-form solution in \cref{sec:closedformHistogram}.
\subsection{Integration with VTK-m and ParaView}\label{sec:vtkmParaview}
Motivated by Wang et al.'s work~\cite{Wang2023}, we integrate the critical point uncertainty code with the VTK-m software~\cite{TA:2016:Moreland:vtkm}. Since the computation of critical point probability depends only on the local neighbors and is independent of data at other pixels, it is embarrassingly parallel. We therefore implemented our code using VTK-m. The advantage of VTK-m is that it allows optimized access to neighbors, and it is platform portable. With our VTK-m implementation, we showcase significant speed-up in critical point uncertainty computation on various architectures, including AMD, NVIDIA, and Intel processors. Further, we export our VTK-m code as a plugin for the use in ParaView software. With our plugin, the uncertainty of critical points can be visualized in ParaView in near-real time and can be combined with the other ParaView filters for better analysis of uncertainty.

\section{Results and Discussions}\label{sec:results}
\subsection{Validation and Performance of Proposed Algorithms}\label{sec:validation}
We validate the correctness of our proposed closed-form computations and algorithms through qualitative and quantitative comparison with respect to the conventional MC sampling approach. We also demonstrate the performance improvements of the proposed methods over MC sampling. \Cref{fig:validationAckley} demonstrates the correctness and enhanced performance of our algorithms through experiments on a synthetic Ackley function~\cite{Ackley1987} sampled on a uniform grid. In particular, the uncertain data is synthetically generated by injecting random uniform noise at each grid position to produce an ensemble of $50$ members. At each grid vertex, the minimum and maximum values are computed from the ensemble to estimate the range of a uniform distribution. The local minimum probability $Pr(p=l_{min})$ and saddle probability $Pr(p=l_{s})$ are then computed at each grid vertex $p$ using the MC sampling method and our proposed algorithms. The results are visualized in \cref{fig:validationAckley}.

Columns \cref{fig:validationAckley}a-b show the results of MC sampling. Column \cref{fig:validationAckley}c shows the results obtained with our closed-form formulation and algorithms. Column \cref{fig:validationAckley}d visualizes the difference image between columns \cref{fig:validationAckley}c and column \cref{fig:validationAckley}b. Column \cref{fig:validationAckley}e visualizes the convergence of MC solutions to our closed-form solution by plotting the root mean squared error (RMSE) between the MC and closed-form solutions. The white isocontours (isovalues $0.1$ and $0.01$ for local minimum and saddle probability results, respectively) shown in \cref{fig:validationAckley}a-c 
 enclose the possible critical point positions. As observed in \cref{fig:validationAckley}a-c, the isocontour structure in MC results converges to our closed-form solutions as we increase the number of MC samples from $100$ to $2000$. 
 
In MC sampling, uniform distributions are sampled to estimate the local minimum/saddle probability. As observed in \cref{fig:validationAckley}a, drawing $100$ samples per grid vertex not only causes more computation time but also results in a lower accuracy compared to our closed-form results in \cref{fig:validationAckley}c. As we increase the number of MC samples to $2000$ in \cref{fig:validationAckley}b, the accuracy increases (i.e., results converge), but the computational performance reduces. Specifically, compared to the MC sampling method with $2000$ samples, our closed-form solution provides a $64 \times$ speed-up for local minimum probability and $119 \times$ speed-up for the saddle probability computation with a comparable accuracy. The timings reported are for the Python serial implementation on a quad-core Intel I7 processor. Note that the MC sampling approach for a saddle takes on average more time compared to MC sampling for the local minimum because of the computation of two terms, $t_1$ and $t_2$, for a saddle (see \cref{sec:criticalPointProbability-FourNeighborhood}). 
The RMSE and maximum probability difference ($Error\_max$) between the MC and our closed-form solutions shown in \cref{fig:validationAckley}d and a convergence curve in \cref{fig:validationAckley}e confirm the correctness of our derivations and algorithms. Similar convergence curves and results for the Epanechnikov and histogram (closed-form and semianalytical) models are reported in the supplement.

\begin{figure}[!b]
  \centering 
\vspace{-4mm}
\includegraphics[width=0.93\linewidth]{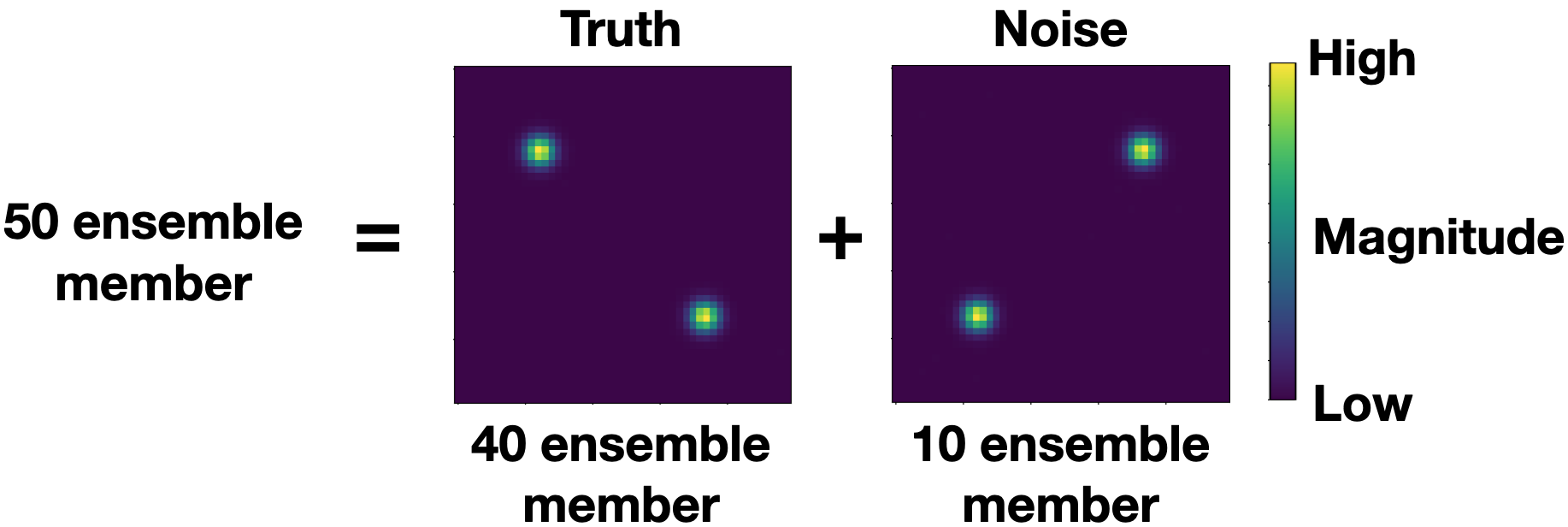}

  \caption{Gaussian mixture data for computation of critical point probability.
  }
  \label{fig:gaussianMixData}
\end{figure}

\begin{figure*}[!t]
  \centering 
  \includegraphics[width=0.97\linewidth]{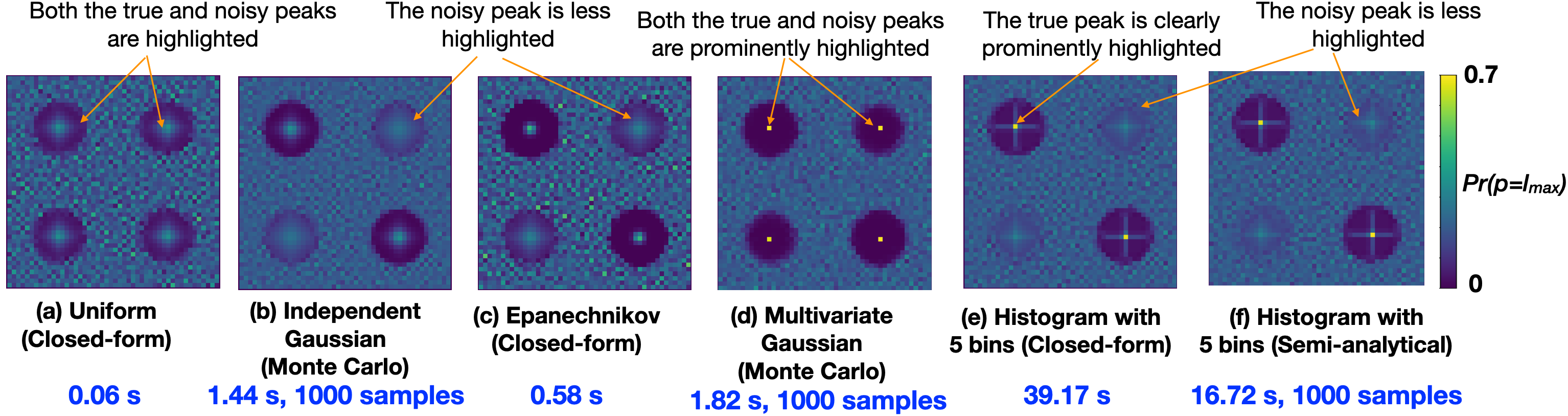}
  \vspace{-2mm}
  \caption{A comparison of parametric vs. nonparametric noise models in terms of quality and performance of local maximum probability (i.e., $Pr[p=l_{max}]$) computation for the ensemble dataset shown in \cref{fig:gaussianMixData}. Our closed-form uniform noise model (a) is the fastest, but it exhibits less accurate results in that it equally highlights both the true and noisy peaks with moderate probability. The independent Gaussian noise (b) model without a closed-form solution and the independent Epanechnikov model with a closed-form solution (c) highlight the true peaks better with respect to the local neighbors, but with moderate probability and slightly more compute time. The multivariate Gaussian model (d) from previous work~\cite{Petz2012criticalPointProbability, Liebmann2016CriticalPointUncertainty} highlights both the true and noisy peaks with high probabilities (bright yellow). The proposed closed-form histogram method (e) exhibits the most robust result to outliers clearly highlighting the true local maximum positions with high probability (bright yellow), but it is the slowest in performance. The proposed semianalytical histogram method (f) reduces the time while maintaining sufficient accuracy when compared to (e).  
  }
  \label{fig:parametricVsNonparametric}
  \vspace{-6mm}
\end{figure*}

\subsection{Comparison of Parametric Vs. Nonparametric Models}\label{sec:modelComparison}

We present a comparison of parametric and nonparametric models in terms of accuracy and performance for computation of critical point probability.
Here, we show the results for the synthetic data modeled as a Gaussian mixture model, which is a common data model used in scientific and topological analysis~\cite{TA:2020:Vidal:persistenceDiagramBarycenter, TA:2020:Yan:mergeTreeAverage,TA:2024:topoSensitivity}. In particular, we generate the ensemble of 50 members, in which each member comprises a mixture of two Gaussians.
As shown in \cref{fig:gaussianMixData}, 40 members correspond to a Gaussian mixture, in which peaks (yellow) are oriented in NW-SE direction. These $40$ members are considered as the ground truth, and the variation across them corresponds to small noise randomly added. The remaining $10$ members are a $90^{\circ}$ rotated version of the first $40$ members oriented in NE-SW direction. These rotated members represent the noisy data or outliers. In other words, the peaks represented by these rotated members are not the true peaks. 

We apply parametric (independent uniform, Epanechnikov, Gaussian, and  multivariate Gaussian) and nonparametric (histogram with $5$ bins) noise models to the Gaussian mixture ensemble (\cref{fig:gaussianMixData}) for visualization of the local maximum probability (i.e., $Pr(p=l_{max})$). The results in \cref{fig:parametricVsNonparametric} demonstrate the quality and performance of various noise models. \Cref{fig:parametricVsNonparametric}a visualizes the result for the independent uniform noise model using our closed-form computational algorithm (\cref{sec:criticalPointProbability-FourNeighborhood}). At each pixel, the data range is computed based on the minimum and maximum data values observed across the ensemble. The closed-form uniform noise model is the fastest, but shows all critical points in NW-SE-NE-SW directions equally likely with moderate probability. This is not a desirable result, as the critical points in the NE-SW directions correspond to noise or outliers (see \cref{fig:gaussianMixData}) and should not result in important features in a probabilistic visualization. 

The quality of results is improved overall in the case of independent Gaussian and Epanechnikov noise models, as observed in \cref{fig:parametricVsNonparametric}b and \cref{fig:parametricVsNonparametric}c, respectively. For both models, we determine the distribution range based on the sample mean and standard deviation per pixel across the ensemble members. Since the Gaussian model does not have a closed-form solution, we performed MC sampling with $1000$ samples to compute the local maximum probability, which is computationally expensive compared to the uniform noise model. On the other hand, the Gaussian noise model highlights the true critical points (NW-SE) better with respect to local neighbors compared to the outlier critical points (NE-SW). However, both true and outlier critical points get moderate probability values assigned. Our proposed closed-form result for the Epanechnikov model in \cref{fig:parametricVsNonparametric}c is similar to the result for the Gaussian model in \cref{fig:parametricVsNonparametric}b, obtained in reduced time with about $2.58 \times$ speed-up. The Epanechnikov model enhances the results compared to the uniform model because of its similar characteristics to the Gaussian model, i.e., greater mean weight, a bell-like shape, and more smoothness, compared to the uniform model (see \cref{sec:parametricModels}). The multivariate Gaussian model in \cref{fig:parametricVsNonparametric}d from the previous work~\cite{Petz2012criticalPointProbability, Liebmann2016CriticalPointUncertainty} prominently highlights both the true and outlier peaks with high probabilities (bright yellow). Capturing data correlation assigns high probability to the outlier critical points and counter intuitively exhibits less robustness to outliers.

\Cref{fig:parametricVsNonparametric}e and \cref{fig:parametricVsNonparametric}f visualize the results for the independent nonparametric histogram model (\cref{sec:nonparametricModels}) with $5$ bins, which exhibit the greatest robustness to outliers. Specifically, they clearly highlight the true local maximum positions with a high probability (bright yellow) and smooth out noisy peaks with a moderate probability (indicated by the arrows in \cref{fig:parametricVsNonparametric}e-f). The closed-form nonparametric models, however, require more computations (see \cref{sec:closedformHistogram}). The computational time of the closed-form nonparametric solution is mitigated with the semianalytical solution (see \cref{sec:semianalytical}) while maintaining comparable accuracy, as seen from \cref{fig:parametricVsNonparametric}e and \cref{fig:parametricVsNonparametric}f. All timing results reported in \cref{fig:parametricVsNonparametric} are again obtained with a serial Python implementation on a quad-core Intel I7 processor. To be able to accommodate the high-quality nonparametric solutions in the visualization systems, we accelerate our algorithms using a C++ parallel implementation with VTK-m~\cite{TA:2016:Moreland:vtkm} (\cref{sec:vtkmParaview}), as presented for the real dataset results in the next section. 

\begin{figure*}[!t]
  \centering 
  \includegraphics[width=\linewidth]{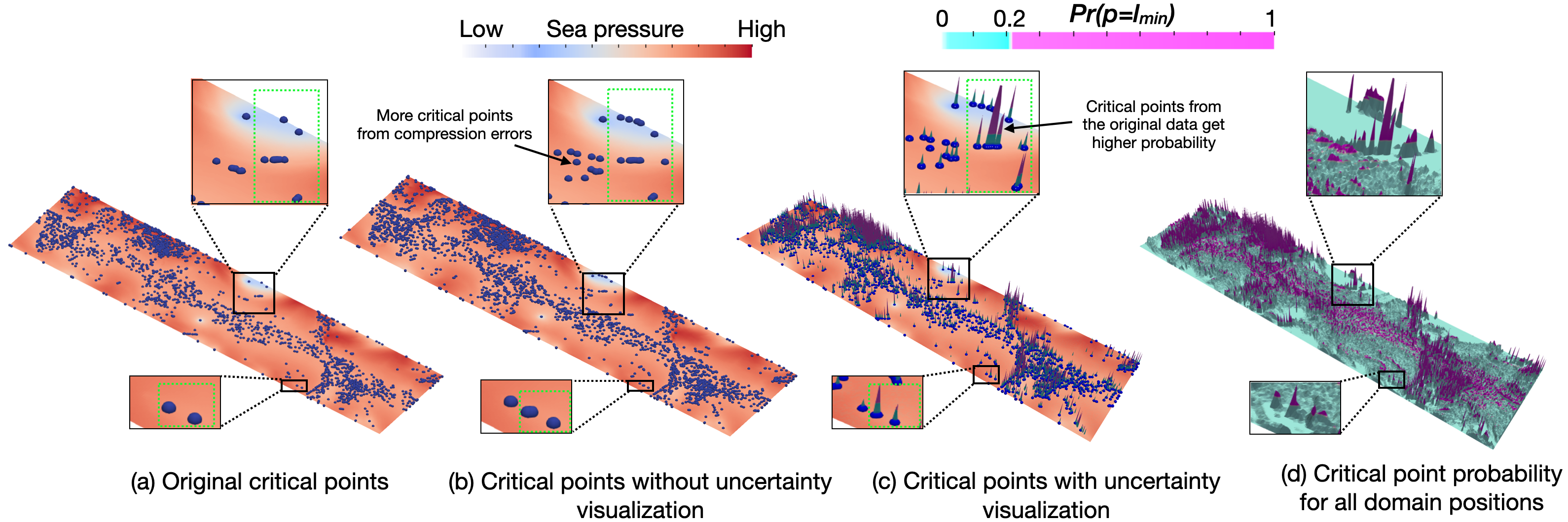}
  \vspace{-5mm}
  \caption{Critical point visualization for the climate dataset. (a) The colormapped original data with local minima shown as the blue spheres. (b) Compression errors result in increased numbers of critical points for which no uncertainty is visualized. (c) Computation of critical point probability with the uniform noise model and visualization of uncertainty through elevation proportional to probability. The critical points in the original data have higher probabilities (tall mountain with magenta color enclosed by the green box). (d) The heightmap of probabilities for every domain position. High probabilities (i.e., tall mountains with magenta peaks) are observed in similar regions as the critical points in the original data.      
  }
  \vspace{-6mm}
  \label{fig:real-climateData}
\end{figure*}

\subsection{Real Datasets}

\paragraph{Climate Data:} \Cref{fig:real-climateData} visualizes the mean sea-level pressure variable simulated from an earth and climate simulation model -- Energy Exascale Earth System Model (E3SM)~\cite{golaz2022doe} -- with a grid resolution of $0.25^\circ$. The result is visualized for a tropical region sampled on a regular grid of size $240 \times 960$, in which the local minima in relatively low pressure regions (i.e., blue) indicate the potential of existence of tropical cyclones. The dataset was compressed using an error-controlled lossy compressor -- MGARD \cite{gong2023mgard} -- under a relative L2 error bound ($eb$) of $\num{1e-3}$, which provided a compression ratio of $16.68$. We study the uncertainty of critical points in data decompressed using MGARD. 

\Cref{fig:real-climateData}a visualizes the original mean sea-level pressure data colormapped with magnitude. The critical points (i.e., local minima $l_{min}$) are shown as the blue spheres extracted using TTK~\cite{TA:2018:TFL}. \Cref{fig:real-climateData}b visualizes the decompressed data and its critical points. As observed, the number of critical points increases in certain regions due to the compression errors, as illustrated by the inset view. There is, however, no indication of how much uncertainty or probability of critical points is in a decompressed field, which can lead to less reliable TDA.
\Cref{fig:real-climateData}c visualizes critical points in a decompressed field along with uncertainty. We utilize the value of error bound $eb$ to derive critical point probabilities. Specifically, at each grid position, the uncertain data range is $[d' - \frac{eb}{2}, d' + \frac{eb}{2}]$, where $d'$ denotes the decompressed value. Since we do not have prior knowledge of the distribution over the uncertain range, we model data uncertainty with the uniform distribution and apply our proposed algorithms (\cref{sec:criticalPointProbability-FourNeighborhood}).  

In \cref{fig:real-climateData}c, we visualize the derived local minimum probability using a heightfield, an idea similar to the previous work by Petz et al.~\cite{Petz2012criticalPointProbability}. In a heightmap, each grid point of interest is elevated and colormapped proportional to the critical point probability. In \cref{fig:real-climateData}c, the heightfield tailored to critical points clearly indicates points with relatively high probability. For the two inset views in \cref{fig:real-climateData}c, the high probability critical points (enclosed by the green dotted box) also appear in the original data in \cref{fig:real-climateData}a. In contrast, newly created critical points due to compression errors have a low probability. \Cref{fig:real-climateData}d visualizes the heightfield for every grid pixel, in which the magenta regions (with $Pr(p=l_{min}) > 0.2$) reflect a pattern of critical point positions that is similar to the one in the original data (\cref{fig:real-climateData}a). We additionally present a quantitative evaluation for the dataset in the supplementary material.  

We measure the performance and accuracy of the climate dataset results through our VTK-m implementation. Since the VTK-m implementation is platform portable, we run it on a serial processor and AMD GPUs on the Oak Ridge National Laboratory's Frontier supercomputer~\cite{Atchley2023} and NVIDIA GPUs on National Energy Research Scientific Computing Center's Perlmutter supercomputer~\cite{Li2023}. Using a conventional MC solution with $1000$ samples per grid point takes $4.94$ seconds on a serial backend. The proposed algorithms for closed-form computation (\cref{sec:criticalPointProbability-FourNeighborhood}) compute the true solution in $0.012$ seconds on a serial backend, thereby providing a $411 \times$ speed-up. Running the VTK-m code on AMD GPU yields a closed-form result in $0.003$ seconds, which corresponds to a $1646 \times$ speed-up compared to the MC solution and a $4 \times$ speed-up compared to the closed-form solution on a serial backend. The NVIDIA GPU yields a closed-form result in $0.004$ seconds, a performance close to the AMD GPU. We present additional accuracy and performance results of our VTK-m code for the AMD GPU in the supplementary material.

\paragraph{Oceanology Data:} In our next experiment, we evaluate the quality and performance of parametric and nonparametric noise models for the Red Sea ensemble simulations~\cite{TA:2020:redSea}. The results are visualized in \cref{fig:teaser}. The dataset is downloaded from the 2020 IEEE SciVis contest website. The ensemble comprises $20$ members each with grid resolution $500 \times 500$. Understanding eddy positions is crucial for oceanologists to gain insight into energy and particle transport patterns in oceans. Therefore, we investigate the local minima positions that potentially correspond to the eddy features. Since the original simulation data are too noisy, we applied topological simplification~\cite{EdelsbrunnerLetscherZomorodian2002} using TTK~\cite{TA:2018:TFL} to each ensemble member as a denoising (preprocessing) step until each ensemble member has approximately the same number of critical points. This strategy corresponds to the {\em persistence graph} idea from the prior work by Athawale et al.~\cite{TA:Athawale2022MsComplex}, which plots the number of local minima against the persistence simplification level to decide the amount of simplification. Having simplified the topology, we efficiently compute and visualize critical point uncertainty in ParaView with our VTK-m code as a backend, as documented in the supplement.

\cref{fig:teaser}a-b visualize the results for the parametric noise models. We apply spherical glyphs in ParaView to help quickly identify the positions with high probability. The points with a larger local minimum probability ($Pr(p=l_{min})$), therefore, have a bigger radius and a red/orange color in the sphere glyphs. The result of the multivariate Gaussian model from previous work by Petz et al.~\cite{Petz2012criticalPointProbability} and Liebmann et al.~\cite{ Liebmann2016CriticalPointUncertainty} with 2000 MC samples is visualized in \cref{fig:teaser}c. Capturing the correlation using the multivariate Gaussian model significantly reduces the probability of local minima in certain regions (blue regions) and emphasizes fewer critical points. \cref{fig:teaser}d visualizes the results for our proposed nonparametric histogram model with four bins (\cref{sec:closedformHistogram}). The uniform, Epanechnikov, multivariate Gaussian, and histogram models took $0.094$, $0.102$, $0.167$ and $0.145$ seconds, respectively, on the Frontier supercomputer's AMD GPU. \cref{fig:teaser}e-f visualize the critical points (yellow) extracted from two random members of the ensemble.

Since we do not know the true critical points for the ensemble data, we make a few interesting observations by comparing results of different noise models (shown with the boxes in \cref{fig:teaser}). The white boxes indicate positions where the two local minimum positions are consistently observed with high probability across all models, and therefore, can be trusted. The green boxes show critical points that are captured as high probability by the proposed nonparametric histogram models, but not by other models. This result is interesting because the histogram showed the highest resilience to outliers, and therefore, more trustworthy results, in our synthetic experiments in \cref{fig:parametricVsNonparametric}. As observed in individual ensemble members, critical points are also seen in the areas marked by the green boxes. Similarly, the cyan boxes mark critical points that are captured by the multivariate Gaussian model, but not by any other models. This result necessitates further investigation because the multivariate Gaussian model was less robust to outliers in our synthetic experiment in \cref{fig:parametricVsNonparametric}. Lastly, the pink boxes show the positions where the multivariate Gaussian and nonparametric models agree, whereas the others do not agree. 
\section{Conclusion and Future Work} \label{sec:conclusion}

In this paper, we study the propagation of uncertainty in critical points, the fundamental topological descriptors of scalar field data. Our main contribution of this paper is the creation of a novel efficient algorithms that compute critical point probability in closed form for parametric and nonparametric uncertainty with finite support. We demonstrate the effectiveness of our algorithms through enhanced accuracy and performance over classical MC sampling. We integrate our algorithms with the VTK-m library~\cite{TA:2016:Moreland:vtkm} to further accelerate performance using serial, AMD, and NVIDIA backends. We show seamless integration of our VTK-m algorithms with ParaView~\cite{TA:2005:paraview} (see the supplement), which is a key to making our algorithms accessible to a wide audience. Our synthetic experiments show the greater resilience of our proposed nonparametric models to outliers compared to parametric models, similar to the prior studies~\cite{TA:Kai:2013:nonparametricIsoVis, TA:Athawale:2016:nonparametricIsosurfaces}. We present the practical utility of our techniques through application to climate and oceanology datasets. 

A few limitations of this work are important and need to be addressed in the future. Currently, we assume that the data at each grid point have uncertainty only over finite bounds. Although the finite bounds assumption is generally true in practice, finding exact bounds that are needed for our proposed algorithms can be nontrivial. In the demonstrated results, we derived the upper and lower bounds for the climate and Red Sea ensemble datasets at each grid point based on the data captured by simulations. Thus, the quality of our results strongly depends on how well the application can provide upper and lower bounds for the values at a grid point. That said, one of the benefits of the proposed nonparametric models is that they can mitigate the effects of overestimation (or coarseness) of bounds resulting from the outliers by assigning a lower weight or probability to outliers (see \cref{sec:nonparametricModels} and the synthetic experiments in \cref{sec:modelComparison}). In the frequent case of ensemble simulations, however, a priori knowledge or experiments demonstrating the robustness of lower and upper bounds provided by the ensembles can help to further improve the trustworthiness of the results.

Another limitation of our work is that we assume  neighboring data points to be uncorrelated. This independent noise assumption can be true in the case of measurement data but is not often true for the real datasets or ensemble simulations encountered in practice~\cite{TA:Pothkow:2011:probMarchingCubes, TA:Athawale:2021:topoMappingUncertaintyMarchingCubes, TA:Brodlie:2012:RUDV, TA:Athawale:2013:lerpUncertainty}. Spatially close grid points are often strongly correlated in real datasets because to reliably represent a smooth function, any reasonable grid must have a spatial resolution finer than frequencies of relevance. Our independent noise assumption, therefore, can lead to overestimation of probabilities from ignoring the local correlation (as shown too in the previous studies~\cite{TA:Pothkow:2011:probMarchingCubes,TA:Athawale:2021:topoMappingUncertaintyMarchingCubes}). Considering spatial correlation, however, has a caveat of higher sensitivity of results to outliers, as demonstrated for the multivariate Gaussian model in our synthetic experiments in \cref{sec:modelComparison}. Even though the core integration formulae in \cref{eq:localMinimumProb} and \cref{eq:saddleProb} are valid for the correlated data, their computation becomes more challenging because of the complexity of accommodating linear or nonlinear correlations. Thus, further research is needed to derive closed-form solutions for critical point probability computation that can accommodate spatial correlation and are robust to outliers.

Our methods are currently limited to critical points of uncertain 2D fields based on four neighbors per grid point. In the future, we would like to extend our work to more neighbors (e.g., six- or eight-pixel neighborhood based on the triangulation) and 3D datasets, which can be complex because of higher order of integration templates. We utilize the sphere glyphs  and heightmaps~\cite{Petz2012criticalPointProbability} for the visualization of critical point probabilities. However, both methods can lead to occlusion and clutter. Thus, further study is needed to evaluate the perceptual quality of sphere glyph and heightmap methods and, possibly, derive new rendering techniques for enhanced perception of uncertainties. Finally, we will also investigate uncertainty in other topological visualizations based on critical points, including contour trees and persistence diagrams.

\acknowledgments{%
	This work was supported in part by the U.S. Department of Energy (DOE) RAPIDS-2 SciDAC project under contract number DE-AC0500OR22725, NSF III-2316496, the Intel OneAPI CoE, and the DOE Ab-initio Visualization for Innovative Science (AIVIS) grant 2428225. This research used resources of the Oak Ridge Leadership Computing Facility (OLCF), which is a DOE Office of Science User Facility supported under Contract DE-AC05-00OR22725, and National Energy Research Scientific Computing Center (NERSC), which is a DOE National User Facility at the Berkeley Lab. We would also like to thank the reviewers of this article for their valuable feedback. %
}

\bibliographystyle{abbrv-doi-hyperref}

\bibliography{main}

\end{document}